# Exploring the Way to Approach the Efficiency Limit of Perovskite Solar Cells by Drift-Diffusion Model

Xingang Ren, Zishuai Wang, Wei E. I. Sha*, Wallace C. H. Choy*

Department of Electrical and Electronic Engineering, The University of Hong Kong, Pokfulam Road, Hong Kong, China

E-mail: chchoy@eee.hku.hk (W.C.H. Choy), wsha@eee.hku.hk (W. E.I. Sha)

## Abstract

Drift-diffusion model is an indispensable modeling tool to understand the carrier dynamics (transport, recombination, and collection) and simulate practical-efficiency of solar cells (SCs) through taking into account various carrier recombination losses existing in multilayered device structures. Exploring the way to predict and approach the SC efficiency limit by using the drift-diffusion model will enable us to gain more physical insights and design guidelines for emerging photovoltaics, particularly perovskite solar cells. Our work finds out that two procedures are the prerequisites for predicting and approaching the SC efficiency limit. Firstly, the intrinsic radiative recombination needs to be corrected after adopting optical designs which will significantly affect the open-circuit voltage at its Shockley–Queisser limit. Through considering a detailed balance between emission and absorption of semiconductor materials at the thermal equilibrium, and the Boltzmann statistics at the non-equilibrium, we offer a different approach to derive the accurate expression of intrinsic radiative recombination with the optical corrections for semiconductor materials. The new expression captures light trapping of the absorbed photons and angular restriction of the emitted photons simultaneously, which are ignored in the traditional Roosbroeck-Shockley expression. Secondly, the contact characteristics of the electrodes need to be carefully engineered to eliminate the charge accumulation and surface recombination at the electrodes. The selective contact or blocking layer incorporated nonselective contact that inhibits the surface recombination at the electrode is another important prerequisite. With the two procedures, the accurate prediction of efficiency limit and precise evaluation of efficiency degradation for perovskite solar cells are attainable by the drift-diffusion model.

Our work is fundamentally and practically important to mathematical modeling and physical understanding of solar cells.

For Table of Contents (TOC)

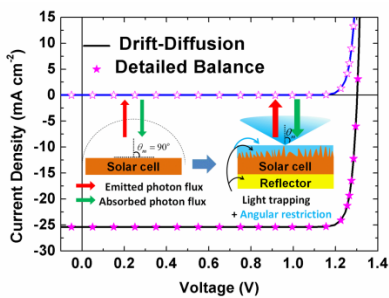

**Introduction**

Owing to the merits of perovskite such as direct bandgap,[1] long carrier diffusion length,[2-4] high charge carrier mobility,[5] small exciton binding energy,[1] and low Auger recombination rate,[6] organic-inorganic perovskite solar cells (PVSCs) have attracted intensive research attention due to their vast potential for addressing energy issues.[7-9] The certified reported power conversion efficiency (PCE) of PVSCs has reached 22.1% from the starting efficiency of 3.9% in 2009.[10, 11] The further improvement of PVSCs efficiency is expected since its predicted efficiency limit is around 31%.[12] Regarding the perovskite materials, the better crystallinity and larger grain size could reduce the losses caused by the unintentional defects and traps for improving the practical efficiency. However, other losses originating from carrier extraction and surface recombination cannot be eliminated even via adopting high-quality perovskite materials. The fundamental working principles of PVSCs should be explored to suppress these losses for approaching the efficiency limit of PVSCs. Moreover, the quantitative analyses of the efficiency degradation caused by different loss mechanisms are of great importance in optimizing PVSC performances prior to the fabrications. The analysis of the PVSC efficiency limit and then offering the way to approach the efficiency limit is paramount for device optimizations. Meanwhile, the calculation of current density-voltage (*J-V*) characteristics and the modeling of carrier dynamics (transport, recombination, and collection) are highly crucial for understanding the device physics of PVSCs.

As a representative theoretical model, detailed balance theory is adopted to predict the efficiency limit of PVSCs, which shows how far the practical efficiency is from its upper limit.[12, 13] However, identifying the loss mechanisms and understanding the loss-induced efficiency degradation are much more important than the efficiency limit itself. It should be noted that detailed balance model is incapable of optimizing PVSC performance and investigating the device physics in carrier dynamics aspects. Differently, the drift-diffusion model enables the optimization of PVSC performances, since the practical device structures of PVSCs are completely incorporated in the model. Moreover, defects,[14] carriers mobility[15] and bulk/surface recombination[16] that pertains to carrier dynamics can be fully depicted through studying drift-diffusion model. Currently, for PVSCs, most of the reported drift-diffusion models [14, 15] are the same as those for inorganic (e.g. silicon) solar cells including the tools of the Solar Cell Capacitance Simulator (SCAPS) [17], Personal Computer One Dimensional (PC-1D) [18], A Device Emulation Program and Tool (ADEPT) [19] and Analysis of Microelectronic and Photonic Structures (AMPS) [20]. However, the prediction of efficiency limit of a solar cell device is still missing in the drift-diffusion model. There are several critical issues such as the connection between the drift-diffusion and detailed balance theory, the accurate determination of the intrinsic radiative recombination with optical corrections for semiconductor materials and the elimination of the external losses (e.g. non-radiative recombination losses, extraction/injection barrier etc.). Although some works have been studying the drift-diffusion model with detailed balance theory, it mainly focused on investigating the photon recycling effects on the device performance improvement. [21, 22] In addition, the revised expression of radiative recombination in the literature is based on the Roosbroeck-Shockley equation which is only suitable for solar cells with a specific light trapping design. [22, 23] For instance, through considering the total internal reflection in semiconductor materials, Marti and co-workers derived the modified Roosbroeck-Shockley expression of the radiative recombination for the planar solar cells in the presence of the back mirror. [23] The expression of radiative recombination is rarely studied for the solar cells with the general light trapping and angular restriction designs. Different from previous literature works, we will explore the way to approach the efficiency limit of PVSCs and then comprehensively understand the device physics for PVSC performance optimizations. To fulfill the objective, we firstly need to explore an approach to derive an exact radiative recombination with the general optical corrections that are capable of capturing the influence of light trapping and angular restriction designs on the device performance of solar cells. Secondly, we should choose the proper boundary conditions at electrodes (i.e. selective and nonselective electrodes) and examine their influences on charge accumulation and surface recombination at the electrodes under the drift-diffusion equation framework.

In this work, through integrating the detailed balance theory with traditional drift-diffusion model, we explore the way for predicting and approaching the efficiency limit of PVSCs and gaining physical understanding of device optimization. At thermal equilibrium, the radiative recombination limit is theoretically derived through the black-body radiation law, which is a different approach as compared to that of the Roosbroeck-Shockley equation. Particularly, the new expression of the radiative recombination takes into account the optical effects such as light trapping and angular restriction, which are ignored in the Roosbroeck-Shockley equation. The proposed approach in the framework of drift-diffusion equations not only accurately predicts the efficiency limit of PVSCs but also well captures the carrier dynamics such as the carrier transport and band bending of quasi-Fermi level. Most importantly, the efficiency degradation originated from different loss mechanisms (e.g. electrode-induced surface recombination and trap-assisted SRH recombination) that can be quantified by the proposed model which is essential to device performance optimizations to approach the Shockley–Queisser limit.

**Formulations**

In this section, we will firstly introduce the fundamentally governing equations for studying the device physics of PVSCs. Secondly, we will theoretically derive the new expression of the radiative recombination with the general optical corrections at the thermal equilibrium for predicting the efficiency limit of PVSCs. Thirdly, the boundary conditions as well as the non-radiative recombination used in the drift-diffusion model will be discussed.

*Governing equations* The device characteristics of PVSCs are governed by the semiconductor equations (including Poisson, drift-diffusion and continuity equations):

$$\nabla \cdot (\varepsilon_r \nabla \psi) = -q(p-n)$$
$$\frac{\partial n}{\partial t} = \frac{1}{q}\nabla \cdot \mathbf{J}_n + G - R \quad (1)$$
$$\frac{\partial p}{\partial t} = -\frac{1}{q}\nabla \cdot \mathbf{J}_p + G - R$$

where $\mathbf{J}_n = -q\mu_n n \nabla \psi + qD_n \nabla n$ and $\mathbf{J}_p = -q\mu_p p \nabla \psi - qD_p \nabla p$ are the electron and hole current densities, respectively. The electron (hole) diffusion coefficient satisfies the Einstein relation $D_{n(p)} = \mu_{n(p)} k_B T/q$ and $\mu_{n(p)}$ is electron (hole) mobility. Furthermore, $G = G^{ph} + G^{dark}$ is the total generation rate, where $G^{ph}$ is the photon generation and $G^{dark}$ is the dark generation at the thermal equilibrium. Similarly, $R=R^{rad}+R^{nonrad}$ is the total recombination at the non-equilibrium in which $R^{rad}$ is the radiative recombination and $R^{nonrad}$ is the non-radiative recombination. Since perovskite is a direct bandgap material, the non-radiative recombination is mainly attributed to defects and traps assisted Shockley-Read-Hall (SRH) recombination and Auger recombination is negligible in PVSCs. [1] In previous work, the bimolecular radiative recombination [24] is utilized to represent the band-to-band recombination of electron-hole pairs with the expression of

$$R^{rad} = B_{rad}(np - n_i p_i)$$
$$B_{rad} = \gamma B_L = \gamma \frac{q}{\varepsilon_0 \varepsilon_r}(\mu_n + \mu_p) \quad (2)$$

where $n_i$ and $p_i$ are the intrinsic densities of electrons and holes, respectively. $B_{rad}$ is the radiative recombination coefficient of PVSCs, which can be empirically represented by a product of the Langevin's recombination coefficient $B_L$ (previously used for organic materials) and reduction factor $\gamma$. Particularly, the radiative recombination of PVSCs is much lower than that of organic solar cells. Through fitting the experimental *J-V* curve of PVSCs, the reduction factor $\gamma$ can be approximately determined and its magnitude is typical on the order of $10^{-4}$. [5, 25] However, the empirical value of the reduction factor $\gamma$ is inconclusive to resolve how far the practical efficiency of PVSCs is from its efficiency limit (as the reduction factor strongly depends on the experimental *J-V* curve). Therefore, the radiative recombination represented by the Langevin model cannot be used to predict the efficiency limit of a solar cell. (See the following results of Figure 1)

*Radiative recombination limit at thermal equilibrium.* To predict the efficiency limit of PVSCs, deriving the radiative recombination limit of perovskite materials at thermal equilibrium is prerequisite. At thermal equilibrium, the dark generation equals to the radiative recombination, that is. $G_0^{dark} = R_0^{rad}$ because of the detailed balance between the emitted photons and absorbed photons. Van Roosbroeck and Shockley previously investigated density of state and the photon-absorbed probability in semiconductor materials and then derived the formulas of the radiative recombination. [26] The Roosbroeck-Shockley expression [21, 26] of the radiative recombination is given by

$$R_0^{rad} = \int_0^\infty \frac{\omega^2 N_r^2 \alpha}{\pi^2 c_0^2} \frac{1}{\exp\left(\frac{\hbar\omega}{k_B T}\right) - 1} d\omega \quad (3)$$

where $N_r$ and $\alpha$ are the refractive index and absorption coefficient of perovskite materials, respectively. [27] Noteworthy, Eq. 3 only relies on the dielectric constant of perovskite. In other words, it is only valid for a homogeneous semiconductor material. As we known, the spontaneous emission rate changes in inhomogeneous electromagnetic environment. [28, 29] Thus, the expression of radiative recombination needs to be corrected especially in the presence of light trapping designs (for enhancing optical absorption) and angular filtering design (for restricting emission angle). [12] The detailed balance between the absorbed photon flux and the emitted photon flux in **Scheme 1a** would be significantly changed after adopting the light trapping designs. Meanwhile, the emission angle can be altered through employing the angular restriction designs as shown in Scheme 1b, which would modulate the flux of the emitted photons and thereby the radiative recombination. In the following section, we will demonstrate that the radiative recombination by the Roosbroeck-Shockley expression underestimate the efficiency limit of PVSCs.

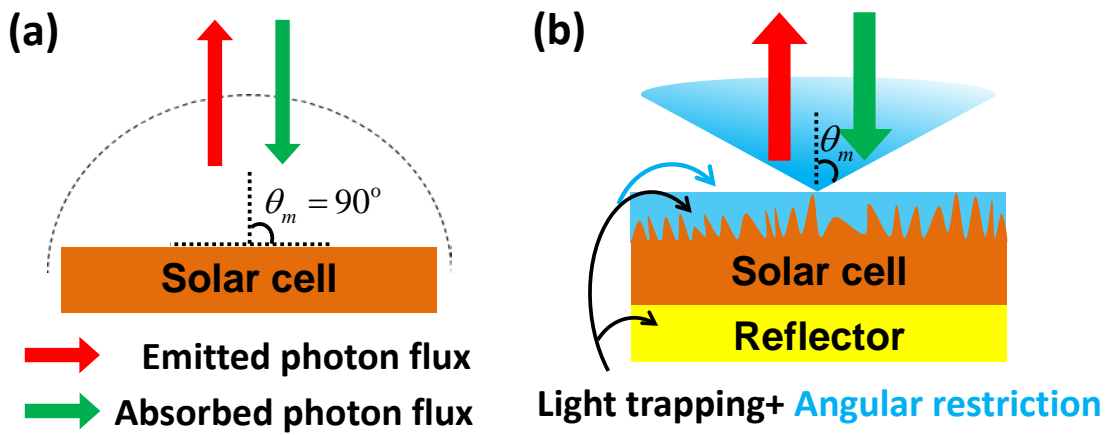

Scheme 1. The emitted and absorbed photon flux at thermal equilibrium for PVSCs with (a) planar structure and (b) light trapping (e.g. back surface reflector and textured front

surface) and angular restriction designs. The maximum emission angle is 90° in (a) and then is restricted to $\theta_m$ in (b) by the incorporation of angular restriction design.

Here, we will propose a different approach to derive a more general expression of the radiative recombination that enable us to take into account the optical manipulations of light trapping and angular filtering designs etc. According to the detailed balance theory and black-body radiation, the dark current can be calculated by the thermal emission that equals to the dark generation at the thermal equilibrium. Using the relationship of $J_0=q\ G_0^{dark}\ L=q\ R_0^{rad}\ L$ ($L$ is the thickness of semiconductor materials), [30] the radiative recombination $R_0^{rad}$ can be retrieved through calculating the dark current $J_0$ which is different from the Roosbroeck and Shockley's work. The dark current itself is proportional to optical absorptivity and black-body emission spectrum of PVSCs. Therefore, in our model, the influences of optical effects on the radiative recombination limit (such as light trapping and angular restriction) have been automatically included during the calculation of dark current. From the black-body radiation law, the dark current of the cell under thermal equilibrium [12, 31] is expressed as

$$J_0 = q\int_0^\infty A(\lambda, L)\frac{\Gamma_0(\lambda)\lambda}{hc_0}d\lambda \qquad (4)$$

in which $A(\lambda, L)$ is the optical absorptivity and $\Gamma_0(\lambda)$ is the black-body (thermal) emission spectrum of the cell. Typically, the absorptivity of a perovskite slab obeys the Beer-Lambert's law. To achieve a desired absorptivity, the perovskite thin film should incorporate a light trapping design (e.g. the randomly textured front surface together with a back mirror reflector). The absorptivity $A$ of PVSCs for achieving the efficiency limit reads [30, 32, 33]

$$A(\lambda, L) = \frac{\alpha(\lambda)}{\alpha(\lambda) + \frac{\sin^2\theta_m}{4N_r^2 L}} \qquad (5)$$

where $L$ is the thickness of perovskite slab, $\theta_m$ is the maximum angle of emission light, that is, light escapes out of the cell within a cone with the solid angle of $\theta_m$. The emission angle $\theta_m$ can be modified through adopting angular restriction filters. The black-body (thermal) emission spectrum of the cell $\Gamma_0(\lambda)$ is expressed by an integral in the spherical coordinate:

$$\Gamma_0(\lambda) = \int_0^{2\pi} d\varphi \int_0^{\theta_m} S(\lambda)\cos(\theta)\sin(\theta)d\theta = \pi \sin^2\theta_m(\lambda) S(\lambda)$$

$$S(\lambda) = \frac{2hc_0^2}{\lambda^5} \frac{1}{\exp\left(\frac{hc_0}{\lambda k_B T}\right) - 1} \quad (6)$$

where $\cos(\theta)$ is due to the angle of emission with respect to the normal of the front surface. $S(\lambda)$ is the thermal spectral radiance (i.e. power emitted per area per wavelength per solid angle) of the cell. After substituting the Eqs. 5 and 6 into Eq. 4 and rewriting the dark current in terms of angular frequency $\omega$, the radiative recombination limit at the thermal equilibrium (without light illumination) is of the form:

$$R_0^{rad} = G_0^{dark} = \int_0^\infty \frac{A(\omega, L)}{N_r^2 L} \frac{\sin^2\theta_m}{4} \frac{\omega^2 N_r^2}{\pi^2 c_0^2} \frac{1}{\exp\left(\frac{\hbar\omega}{k_B T}\right) - 1} d\omega \quad (7)$$

After applying the external bias and illuminating with sunlight, the solar cell works at the non-equilibrium state. In this situation, the radiative recombination obeys the Boltzmann statistics and will exponentially increase with $\exp(\Delta E_F/k_B T)$, where $\Delta E_F = E_{Fn} - E_{Fp}$ is the quasi-Fermi level splitting, $E_{Fn} = E_C + k_B T \ln(n/N_C)$ and $E_{Fp} = E_V - k_B T \ln(p/N_V)$ are the quasi-Fermi level of the electrons and holes, respectively. [34] Finally, Eq. 7 is revised as:

$$\tilde{R}_0^{rad} = \tilde{G}_0^{dark} = \int_0^\infty \frac{A(\omega, L)}{N_r^2 L} \frac{\sin^2\theta_m}{4} \frac{\omega^2 N_r^2}{\pi^2 c_0^2} \frac{1}{\exp\left(\frac{\hbar\omega - \Delta E_F}{k_B T}\right) - 1} d\omega \quad (8)$$

We have offered a different approach to derive the radiative recombination (Eq. 8) that is the intrinsic radiative recombination for semiconductor materials with the optical corrections. Interestingly, under the weakly absorbing limit ($\alpha L \ll 1$), the absorptivity $A$ can be expressed by an approximate form of the Beer-Lambert's law ($A \approx \alpha L$), the radiative recombination limit of Eq. 7 is reduced as:

$$R_0^{rad} = G_0^{dark} = \int_0^\infty \frac{\sin^2 \theta_m}{4N_r^2} \frac{\omega^2 N_r^2 \alpha}{\pi^2 c_0^2} \frac{1}{\exp\left(\frac{\hbar\omega}{k_B T}\right) - 1} d\omega \tag{9}$$

Compared to the Roosbroeck-Shockley equation (Eq. 3), Eq. 9 for the radiative recombination limit includes two additional factors of $\sin^2(\theta_m)$ and $4N_r^2$, which relate to the maximum emission angle and absorption limit, respectively. Eq. 9 indicates that the optical designs will modify the radiative recombination limit. For more rigorous calculations, Eq. 8 is the general formula of the radiative recombination limit where the effects of light trapping and angular restriction are included.

*Non-radiative SRH recombination.* In a practical device, the non-radiative recombination plays an important role to degrade the efficiency, which is a key loss factor that must be minimized for approaching the efficiency limit. To gain the insight of the device physics and optimizations, it is essential to study the recombination losses in PVSCs. The non-radiative recombination in PVSCs is mainly due to the following trap-assisted SRH recombination, [16, 24] that is,

$$R^{nonrad} = R_{SRH} = B_{SRH}(np - n_i p_i)$$
$$B_{SRH} = \frac{C_n C_p N_t}{C_n(n + n_1) + C_p(p + p_1)} \tag{10}$$

where $C_n$ and $C_p$ are the capture coefficients for electrons and holes at the trapped sites, respectively. $N_t$ is the density of trap sites with the trap level $E_t$ in the bandgap, and $n_1 = N_C \exp[(E_t - E_C)/k_B T]$ and $p_1 = N_V \exp[(E_V - E_t)/k_B T]$ are the densities of electrons and holes, respectively.

*Boundary conditions.* Mathematically, the boundary conditions are vital to the final solutions of the semiconductor equation, while physically they will induce the charge accumulation and surface recombination and prevent the practical PVSC from achieving its efficiency limit. The boundary conditions of the electrodes are represented by the surface recombination velocities. The expressions are of the following forms:

$$cathode\begin{cases} J_p^{cat} = qS_p^{cat}(p - p_{eq}^{cat}) \\ J_n^{cat} = qS_n^{cat}(n - n_{eq}^{cat}) \end{cases} \quad anode\begin{cases} J_p^{an} = qS_p^{an}(p - p_{eq}^{an}) \\ J_n^{an} = qS_n^{an}(n - n_{eq}^{an}) \end{cases}$$

$$cathode\begin{cases} n_{eq}^{cat} = N_c \\ p_{eq}^{cat} = N_v \exp\left(\frac{-E_g}{k_B T}\right) \end{cases} \quad anode\begin{cases} p_{eq}^{an} = N_v \\ n_{eq}^{an} = N_c \exp\left(\frac{-E_g}{k_B T}\right) \end{cases} \quad (11)$$

where $S_n^{cat}$ and $S_n^{an}$ ($S_p^{cat}$ and $S_p^{an}$) are the surface recombination velocities of electrons (holes) at the cathode and anode, respectively. $n_{eq}$ and $p_{eq}$ are the electron and hole densities under the assumption of infinite surface recombination velocities. For simplicity, in the following context, we consider the surface recombination velocities for the majority/minority charge carrier are symmetric that is, $S_n^{cat} = S_p^{an}$ and $S_n^{an} = S_p^{cat}$. The cathode is defined as $x=0$ and anode is $x=L$. The selective contacts (i.e. $S_n^{cat} = S_p^{an} = \infty$ and $S_n^{an} = S_p^{cat} = 0$) indicate that there is no surface recombination of minority charge carriers (i.e. no surface recombination of electrons at the anode and no surface recombination of holes at the cathode). The nonselective contacts are referred as $S_n^{cat} = S_p^{an} = S_n^{an} = S_p^{cat} = \infty$ meaning that the minority charge carriers will recombine at the electrodes. In other words, the minority electrons (holes) that diffuse to anode (cathode) will recombine with the thermally activated holes (electrons) at the anode (cathode). In the following contents, the mentioned surface recombination in PVSCs specifically indicates the recombination around the contact interfaces.

**Results and Discussions**

In this section, we conduct the numerical studies to demonstrate the capability of the proposed drift-diffusion model for predicting the typical $CH_3NH_3PbI_3$ (with bandgap of 1.5 eV) based PVSC efficiency limit and evaluating the efficiency degradation induced by the trap-assisted SRH recombination. We will firstly investigate device performances of PVSCs with nonselective and selective contacts by studying the spatially distributed charge carriers. Secondly, we will offer the approach to accurately predict the efficiency limit of PVSCs by the drift-diffusion model. The last part will evaluate the PVSC efficiency degraded by the trap-assisted SRH recombination.

In the simulation, the Poisson and drift-diffusion equations are solved by the finite-difference method with an explicit scheme in space domain and semi-implicit scheme in time domain. [35] The studied planar device structure is (selective and nonselective) cathode/perovskite layer/ (selective and nonselective) anode. The nonselective contacts are adopted to consider the existence of surface recombination at the electrodes, the influences of the minority carriers at the imperfect contacts on device performance can be studied by the nonselective contacts with different surface recombination velocities. In the prediction of the PVSC efficiency limit, the selective contacts of cathode and anode are adopted and they function as the "perfect" blocking layers that would inhibit the surface recombination at the electrodes. The studies of the selective and nonselective contact provide us the comprehensively physical understanding of the loss mechanisms in PVSCs. It should be noted that in this work the recombination losses studied in the PVSCs include surface recombination at the electrodes, radiative and non-radiative (SRH) recombination in the bulk. The detailed parameters of $CH_3NH_3PbI_3$ perovskite materials used in the simulation are listed in **Table S1**.

*Selective and nonselective contacts.* We firstly investigate the *J-V* characteristics of PVSCs with the selective and nonselective contacts. Instead of using the newly derived expression of intrinsic radiative recombination (Eq. 8), the bulk radiative recombination expressed by the Langevin's recombination coefficient $B_L$ and reduction factor $γ$ (Eq. 2) is adopted to demonstrate the impacts of selective/nonselective contacts and radiative recombination on device performance. The non-radiative (SRH) recombination in the bulk is not included in this subsection (SRH recombination will be discussed in the following subsection of *Trap-assisted SRH Recombination*), and the involved recombination losses in the PVSCs only consist of the radiative recombination in bulk and surface recombination at electrodes. As shown in **Figure 1**, when the empirical reduction factor $γ$ decreases, the open circuit voltage ($V_{OC}$) continually increases for PVSCs under the selective contacts condition. Differently, for PVSCs with the nonselective contacts, $V_{OC}$ becomes saturated when the value of the reduction factor $γ$ is less than $10^{-2}$ (**Figure 1b**). As we know, the magnitude of the radiative recombination of PVSCs is typically four-orders of magnitude lower than that of Langevin's recombination, i.e. $\gamma \approx 10^{-4}$.[5] Therefore, the existence of the surface recombination induced by the nonselective contacts sets a fundamental limit to achieve high-efficiency PVSCs.

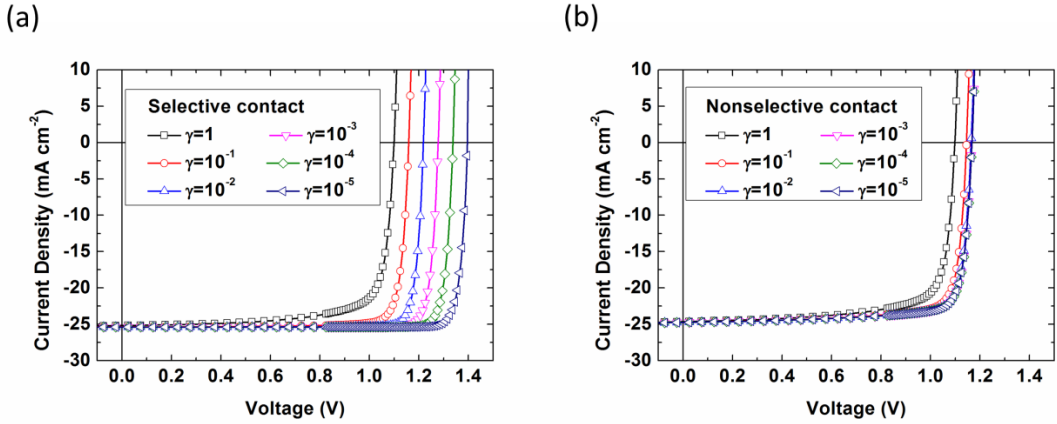

Figure 1. The current density-voltage (*J-V*) characteristics of PVSCs with (a) selective and (b) nonselective contacts. The radiative recombination is modified by changing the reduction factor of *γ*.

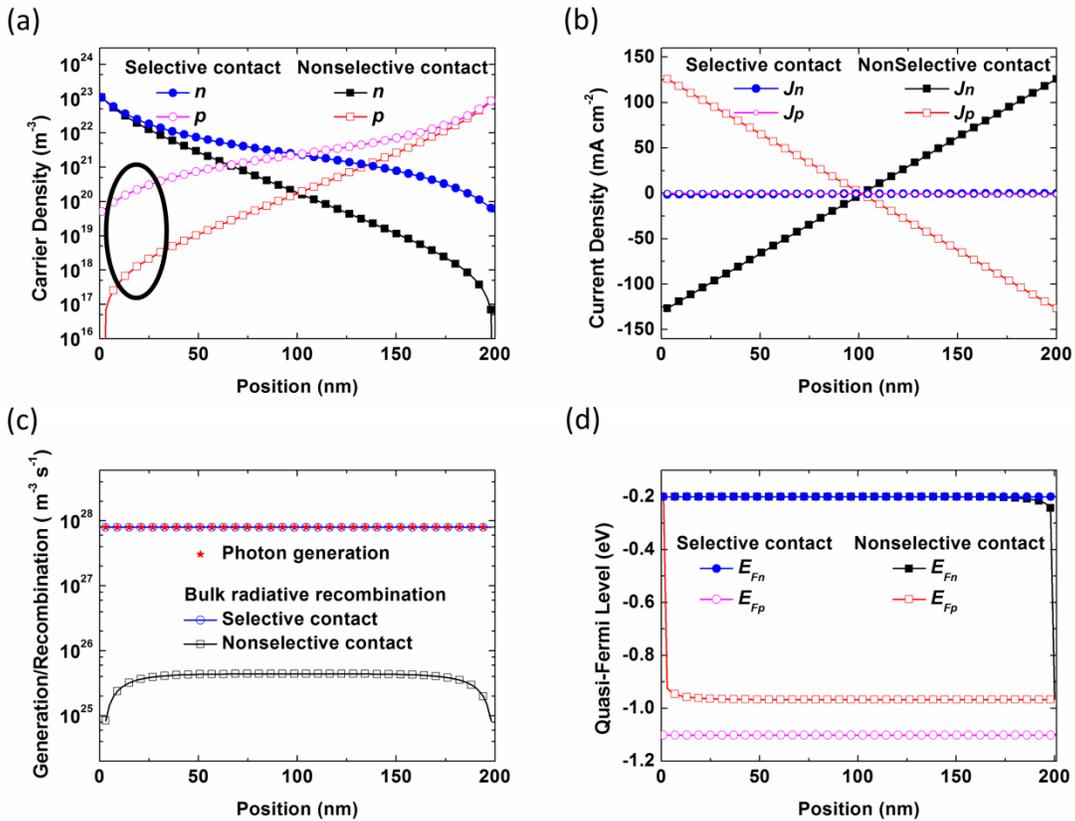

Figure 2. Carrier dynamics in the active layer of PVSCs with the selective and nonselective contacts under the $V_{OC}$ condition. The reduction factor is set to $\gamma =10^{-4}$. (a) Carrier density (*n*, *p*), the circle denotes the hole density near the cathode in PVSCs with selective and nonselective contacts; (b) Current density ($J_n$, $J_p$); (c) Bulk (radiative) recombination and photon generation; (d) Quasi-Fermi level, $E_{Fn}$ and $E_{Fp}$ are the quasi-Fermi levels for electrons and holes, respectively.

The PVSC performances can also be understood by the spatially distributed charge carriers under the $V_{OC}$ condition. As shown in **Figure 2a**, the density of holes at the cathode (*x*=0) with the nonselective contacts (open-red rectangular) is smaller than that of the case of the selective contacts (open-purple circle). Regarding the nonselective contacts, the minority holes that diffuse to the cathode will recombine with the thermally activated electrons at the cathode under the condition of an infinite surface recombination velocity (i.e. $S_p^{cat}=\infty$), leading to a giant recombination loss. On the other hand, the selective contacts suppress the giant carrier recombination and thus possess large charge carrier densities. For the current density, it is well known that the total current density (**J**=**J**$_n$+**J**$_p$) should be zero across the whole device under $V_{OC}$ condition. As shown in **Figure 2b**, the current density of both electrons and holes are exactly zero (**J**$_n$=**J**$_p$=0) in the whole device region under the selective contact condition. Contrarily, with the nonselective contacts, the electrons and holes current density is non-zero despite of null net current.

As shown in **Figure 2c**, with the nonselective contacts (induce surface recombination at electrodes), the bulk (radiative) recombination is 2 orders of magnitude smaller than photon generation. Since the total (i.e. the summation of surface and bulk) recombination must equal to the photon generation at $V_{OC}$, it indicates that the surface recombination at the nonselective electrodes is the dominated recombination loss as compared to the bulk radiative recombination. Considering the selective contact case, a high $V_{OC}$ is achieved when the bulk radiative recombination equals to the photon generation, confirming the absence of the surface recombination loss at the selective contacts. Another interesting thing is the band bending that exists in the PVSCs with the nonselective contact. As seen in **Figure 2d**, the band bending of the quasi-Fermi levels of electrons and holes exists close to the electrodes. The space-charge accumulation of carriers near the electrodes is an in-depth reason for the band bending. The accumulated charge carriers at the electrodes also lead to a large potential change in the nonselective contact case, as compared to that of the selective contacts (See **Figure S1a**). Therefore, the proposed drift-diffusion model enables us to study the selective and nonselective contacts in the PVSCs from the carrier dynamics aspects. With the physical understanding of Langevin's type of radiative recombination and the spatially distributed carrier density under selective and nonselective contacts, we will show that the selective contact and accurate expression of radiative recombination are of critical importance for predicting and approaching the PVSC efficiency limit.

*Prediction of PVSC efficiency limit.* In the theoretical discussions at the subsection of *Selective and nonselective contacts*, the radiative recombination ("intrinsic") in the drift-diffusion model affects the PVSC efficiency strongly since there is no "extrinsic" loss mechanism (e.g. surface recombination at electrodes, trap-assisted non-radiative recombination etc.) in the PVSCs with the selective contacts. Although the employment of the Langevin's type of radiative recombination (the formula is previously used for organic materials) can understand PVSC performances, it cannot be used to predict the efficiency limit. As shown in **Figure S1b**, the predicted $V_{OC}$ of PVSCs could go beyond the $V_{OC}$ limit when the reduction factor $\gamma$ in Langevin's model is less than $10^{-5}$. To accurately predict the efficiency limit of PVSCs, the newly derived expression of the radiative recombination (Eq. 8) should be incorporated into the drift-diffusion model in which the quasi-Fermi level splitting $\Delta E_F$ approximately equals to the applied bias. The *J-V* characteristics of PVSCs with the selective and nonselective contacts are shown in **Figure 3a**. The summary of photovoltaics parameters are listed in **Table 1**. For the PVSCs with the nonselective contacts, the existence of surface recombination of minority carriers at the electrodes is detrimental to PVSC performances. The small $V_{OC}$ of 1.17 V together with a lower fill factor (FF) and short-circuit current density ($J_{SC}$) results in a lower PCE of 23.83%. In contrast, PCE of the PVSCs with the selective contacts reaches to 29.87% due to the high $V_{OC}$ of 1.30 V as well as the improved FF and $J_{SC}$. The existence of surface recombination at the electrodes would degrade the PVSC efficiency from the efficiency limit of 29.87% to 23.83% with a decreasing rate of 20.20%. Moreover, it is noteworthy that the predicted parameters of *J-V* characteristics by the drift-diffusion model are consistent with the results predicted by the detailed balance model (See **Table 1** and **Figure S2a**).[12] However, when the Roosbroeck-Shockley expression (Eq. 3) of radiative recombination is employed (See **Figure S2b**), the predicted $V_{OC}$ of the PVSCs with the selective contacts is reduced to 1.24 V (that corresponds to PCE of 28.39%). Because the radiative recombination by the Roosbroeck-Shockley expression is only valid in homogeneous semiconductor materials, the incorporation of light trapping and angular restriction designs would change the local density of states in perovskite materials that would modify the absorptivity/emission of perovskite thereby the radiative recombination (Eq. 8). As shown in **Figure S3**, the $V_{OC}$ further increases from 1.30 V to 1.34 V through restricting the emission angle from 90$^o$ to 30$^o$ that corresponds to the increase of PVSC efficiency limit from 29.87% to 30.76%. Regarding the prediction of efficiency limit, the $V_{OC}$ predicted by the Roosbroeck-Shockley expression (1.24 V lead to PCE=28.39%) is lowered by 0.10 V (correspondingly underestimate PCE of 2.37%) without considering the light trapping and angular restriction designs. Consequently, the efficiency limit of PVSCs will be underestimated without properly considering the optical effects. To approach the efficiency limit with the drift-diffusion model, the utilization of the newly derived radiative recombination together with the selective contacts is the perquisite..

Meanwhile, it can be seen that the dark *J-V* curves predicted by the drift-diffusion model and detailed balance model are also consistent. (See **Figure 3b**) Interestingly, by simply shifting the simulated dark *J-V* curve (by the drift-diffusion model with selective contacts) down with the value of $J_{SC}$ (25.38 mA cm$^{-2}$ from **Table 1**), we find that the shifted dark *J-V* curve is identical to the simulated *J-V* curve under light illumination (denoted as the photon *J-V*, see **Figure 3b**). The overlap of the downward shifted dark *J-V* curve and photon *J-V* curve indicates that the light illumination dependent losses are excluded in the drift-diffusion model. In other words, during the prediction of PVSC efficiency limit, only the radiative recombination exists, other losses including the trap-assisted SRH recombination, electrode-induced surface recombination, etc. are entirely eliminated in the model. Therefore, starting from the PVSCs with only radiative recombination (i.e. no extrinsic loss mechanism), the drift-diffusion model developed can quantify the efficiency degradation by different loss mechanisms term by term through incorporating the SRH recombination, surface recombination and injection/extraction barrier into the model, respectively. The drift-diffusion model allows us to gain more physical insights of device optimizations.

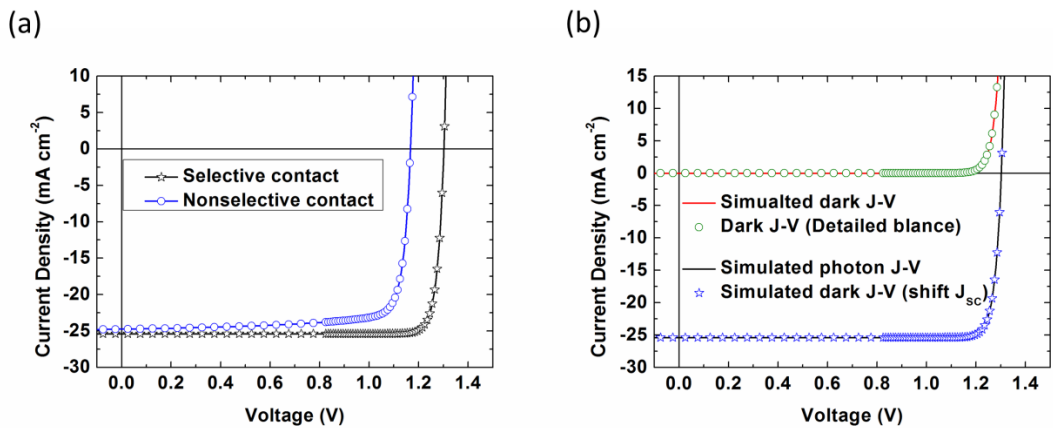

Figure 3. The *J-V* characteristics of PVSCs. (a) The simulated *J-V* curves with the selective and nonselective contacts by the drift-diffusion model. (b) The dark *J-V* curves are obtained by the drift-diffusion model and detailed balance model, respectively. The downward shifted dark *J-V* curve with the value of $J_{SC}$ (=25.38 mA cm$^{-2}$) is identical to the simulated *J-V* curve under light illumination (denoted as the photon *J-V*).

**Table 1**. The summary of PVSC performances.

| Contact | $V_{OC}$ (V) | $J_{SC}$ (mA cm$^{-2}$) | FF (%) | PCE (%) |
|---|---|---|---|---|
| Nonselective contact | 1.17 | 24.74 | 82.52 | 23.83 |
| Selective contact | 1.30 | 25.38 | 90.39 | 29.87 |
| Detailed balance model | 1.30 | 25.38 | 90.39 | 29.86 |

In a practical PVSC, the existence of minority charge carriers at the two electrodes (due to carrier diffusion) will induce a significant surface recombination at the imperfect contacts. As shown in **Figure S4**, for the nonselective electrode case, the increase of the surface recombination velocity from 0 to $10^4$ cm s$^{-1}$ for minority charge carriers significantly reduces $V_{OC}$ from 1.30 to 1.17 V. Meanwhile, the existence of surface recombination of minority carrier would induce the space-charge accumulation and therefore is harmful to the carrier extraction resulting in a reduction of FF from 90.39 to 82.52. Differently, $J_{SC}$ only slightly decreases due to a large built-in potential and thus ignorable recombination. Here, we will show that the surface recombination of minority charge carriers can be totally prohibited in the case of nonselective contact through introducing a blocking layer inserted between the electrodes and active layer (**See Figure S5**). As shown in **Figure 4a**, when the energy barrier of the blocking layers increases to 0.21 eV, performance of the PVSCs with the nonselective contacts and blocking layers becomes identical to that of the PVSCs with the selective contacts. Consequently, the insertion of the blocking layers improves PVSC efficiency approaching to the efficiency limit because of the eliminated surface recombination at the electrodes. Additionally, the PVSC performance cannot be significantly improved by only inserting one blocking layer (See **Figure 4b**). The PVSC efficiency is limited by the surface recombination at the other contact without the blocking layer, which has a large surface recombination. In conclusion, the reduction of the surface recombination of minority carriers at both the cathode and anode is of equal importance. The selective contacts are the essential constituents for predicting the PVSC efficiency limit.

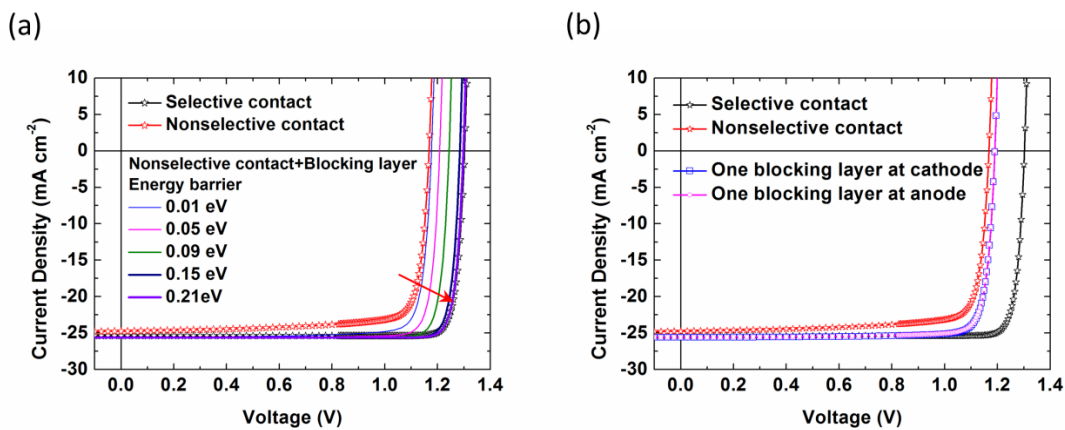

Figure 4. The *J-V* characteristics of PVSCs integrating the nonselective contacts with the blocking layers with different potential barriers. (a) Blocking layers with the same energy barriers; (b) PVSCs with one or two blocking layers. The arrow direction in (a) indicates the increase of the energy barrier from 0.01 to 0.21 eV.

*Trap-assisted SRH recombination.* Since perovskite is a direct bandgap material, the non-radiative recombination is mainly due to the trap-assisted SRH recombination. The density of the trap states is mainly determined by the amount of defects due to the low perovskite crystallinity. The reported trap density ($N_t$) of high-quality film is below $10^{15}$ cm$^{-3}$. [1, 36, 37] The energy level of the trap sites $E_t$ is typically shallow and is around 0.10 eV. [1] As shown in **Figure 5a**, for the PVSCs with the selective contacts, the predicted efficiency is approaching to the efficiency limit of PVSCs when the trap density decreases to $N_t = 10^{15}$ cm$^{-3}$ (the trap energy level is considered to be $E_t=0.10$ eV). In other words, further improvement of perovskite crystallinity (i.e. reduction of trap density) cannot provide a significant enhancement of PVSC efficiency, which is consistent to the experimental observation. The results also reveal that the $V_{OC}$ continuously decreases from 1.30 V to 1.18 V as the trap density $N_t$ increases from 0 to $10^{17}$ cm$^{-3}$. The corresponding PCE is reduced from 29.87% to 26.51% (See **Table S2**). The increment of the trap energy level to 0.15 eV would further decrease the $V_{OC}$ to 1.14 V. This corresponds to a degradation of PVSC efficiency from its efficiency limit of 29.87% to the practical efficiency of 24.58% with a reduction rate of 17.70% (See the inset of **Figure 5a** and **Table S2**). Differently, for PVSCs with nonselective contacts, there is no significant decrease of $V_{OC}$ even the trap density in PVSCs is up to $10^{17}$ cm$^{-3}$ (See **Figure 5b**). Our model enables us to understand the distinct dependence of *J-V* characteristics on the trap density since there are only two loss mechanisms included in the drift-diffusion model. For PVSCs with nonselective contacts, there is a competition between the loss mechanisms of the surface recombination at the electrodes and SRH recombination in the bulk. Regarding the PVSCs in the presence of a low trap density ($<10^{17}$ cm$^{-3}$), the efficiency degradation induced by the SRH recombination in the bulk is marginal as compared to surface recombination at electrodes due to the high charge carrier mobilities.

When the trap density increases up to $10^{19}$ cm$^{-3}$, the charge carrier can be readily trapped by the large amount of the defects before the drift/diffusion of the charge carriers to the respective electrodes, thus the SRH recombination in the bulk becomes the dominant loss mechanisms. Consequently, the drift-diffusion model developed not only quantitatively evaluates the efficiency degradation caused by different loss mechanisms but also provides physical insight into device performance optimization.

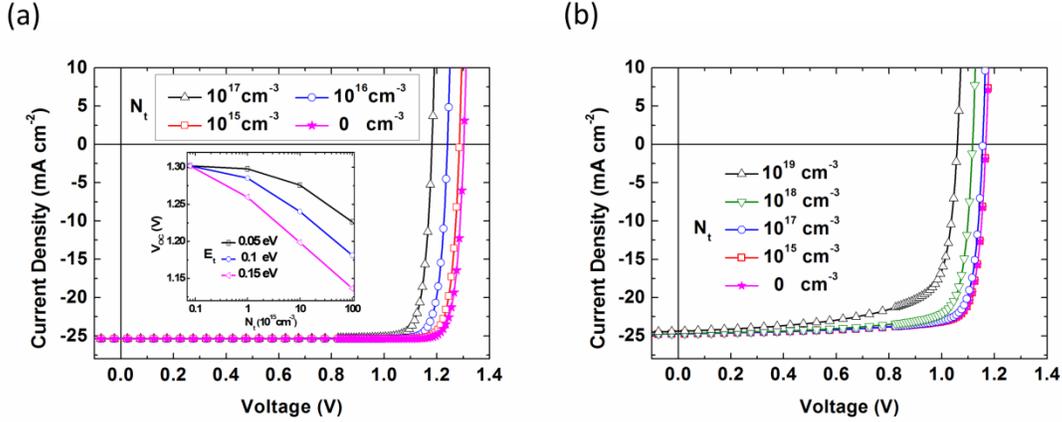

Figure 5. The *J-V* characteristics of PVSCs with different densities of trap states $N_t$ at the trap energy level of $E_t$=0.10 eV. The adopted electrodes in calculation are selective contacts in (a) and nonselective contacts in (b). The inset in (a) shows the relation between the $V_{OC}$ and density of trap states $N_t$ under the trap energy level of $E_t$=0.05, 0.10 and 0.15 eV.

**Conclusion**

In conclusion, under the drift-diffusion equation framework, we have explored and offered the way to approach the efficiency limit and have examined the practical device physics of PVSCs. The optically corrected radiative recombination and the selective contact are two essential prerequisites that enable to approach the efficiency limit by the drift-diffusion model. Through taking account of the detailed balance theory and black-body radiation, the radiative recombination limit has been theoretically derived in a different approach as compared the traditional Roosbroek-Shockley expression. The optical effects involving light trapping and angular restriction were included in terms of the absorptivity and emission angle, respectively. More importantly, the PVSC efficiency degraded by different loss mechanisms can be quantitatively evaluated by the drift-diffusion model. The degradations of PVSC efficiency due to the surface and trap-assisted SRH recombination were discussed. Consequently, by the drift-diffusion model developed, we can not only accurately predict the PVSC efficiency which is highly important to optimize the device performance to approach its efficiency limit, but also offer a comprehensively physical understanding of the losses in the PVSCs through quantifying the efficiency degradation induced by different loss mechanisms term by term.

**Acknowledgment**

This project was supported by the General Research Fund (Grants HKU711813 and 17211916), the Collaborative Research Fund (Grants C7045-14E) from the Research Grants Council of Hong Kong Special Administrative Region, China, ECF Project 33/2015 from Environment and Conservation Fund, and Grant CAS14601 from CAS-Croucher Funding Scheme for Joint Laboratories.

Supporting Information Available: Parameters used in the modeling, current density-voltage curves of perovskite solar cells with different contacts and expressions of radiative recombination, schematic illustration of the insertion of blocking layer, etc. This material is available free of charge via the Internet at http://pubs.acs.org

This work has been published in *ACS Photonics*, 2017, 4 (4), pp 934–942

**Supporting Information for**

**Exploring the Way to Approach the Efficiency Limit of Perovskite Solar Cells by Drift-Diffusion Model**

Xingang Ren, Zishuai Wang, Wei E. I. Sha*, Wallace C. H. Choy*

Department of Electrical and Electronic Engineering, The University of Hong Kong, Pokfulam Road, Hong Kong, China.

E-mail: chchoy@eee.hku.hk (W.C.H. Choy), wsha@eee.hku.hk (W. E.I. Sha)

Table S1. The physical parameters used in the drift-diffusion model.

| Name | Symbol | Numerical Value |
|---|---|---|
| Bandgap | $E_g$ | 1.5 eV [1] |
| Conduction/Valence band | $E_C/E_V$ | -3.9/-5.4 eV [1] |
| Electron and hole mobility | $\mu_n, \mu_p$ | 10 cm$^{-3}$ s$^{-1}$ [2-4] |
| Density of state of conduction and valence band | $N_C, N_V$ | $10^{21}$ cm$^{-3}$ [4, 5] |
| Thickness of photon active layer | $L$ | 200 nm |
| Photon generation | $G$ | $8 \times 10^{21}$ cm$^{-3}$ s$^{-1}$ |
| Radiative recombination limit | $R_0^{rad}(=G_0^{dark})$ | $1.07 \times 10^6$ cm$^{-3}$ s$^{-1}$ |
| Dielectric constant | $\varepsilon_r$ | 10 [4, 5] |
| Room temperature thermal energy | $k_B T$ | 25.7 meV |
| Maximum emission angle | $\theta_m$ | 90º |
| Capture coefficient for electrons at trap states | $C_n$ | $10^{-8}$ cm$^3$ s$^{-1}$ [6] |
| Capture coefficient for holes at trap states | $C_p$ | $10^{-5}$ cm$^3$ s$^{-1}$ [6] |
| Density of trap states | $N_t$ | $10^{15} \sim 10^{17}$ cm$^{-3}$ [6, 7] |
| Energy level of trap sites | $E_t$ | 0.05, 0.10, 0.15 eV |

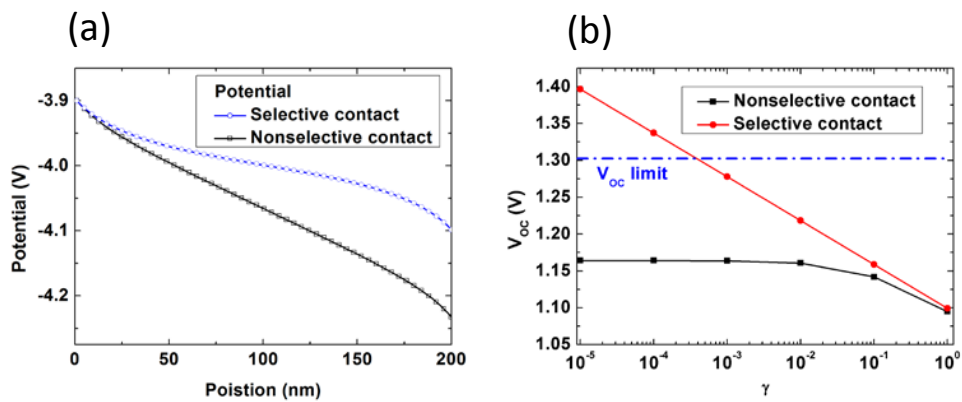

figure S1. (a) Electrical potential in the PVSCs with the selective and nonselective contacts, respectively. (b) $V_{OC}$ of the PVSCs with the selective and nonselective contacts using different reduction factor $\gamma$. The dash-dot line represents the $V_{OC}$ limit (from Table 1).

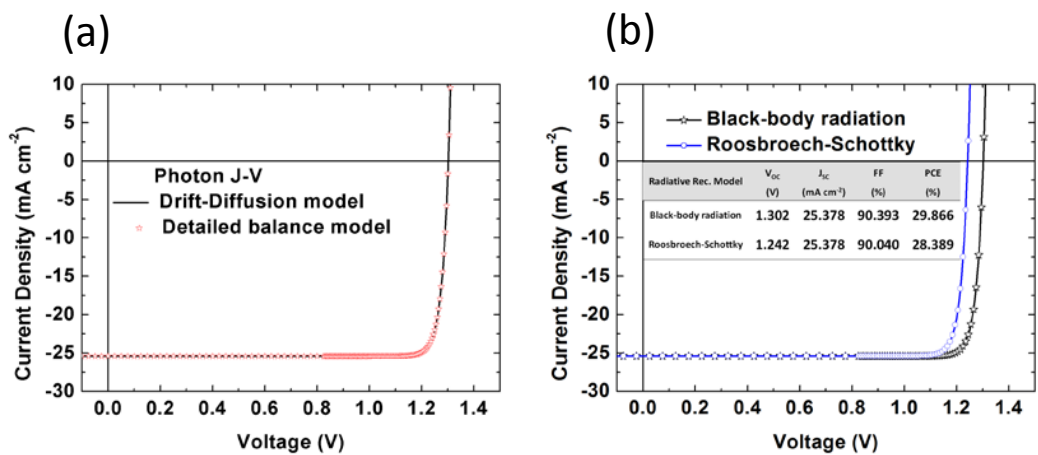

Figure S2. The *J-V* characteristics of PVSCs. (a) Comparisons of *J-V* curves under the light illumination (denoted as photon *J-V*) obtained by the drift-diffusion and detailed balance model, respectively. (b) Comparisons of *J-V* characteristics with the radiative recombination determined by the black-body radiation (Eq. 8) and the Roosbroech-Schottky expression (Eq. 3), respectively. The inset is the corresponding *J-V* characteristics of PVSCs.

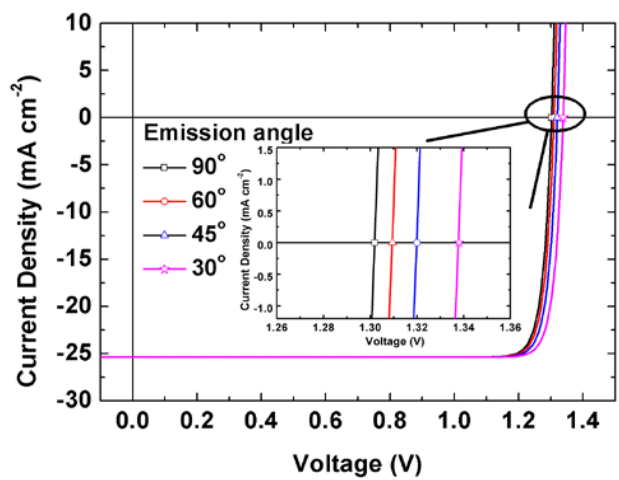

Figure S3. The *J-V* characteristics of PVSCs with different emission angles. The PVSCs under investigation is with selective contact and there is no trap-assisted non-radiative (SRH) recombination in the perovskite. The decrease of the emission angle from 90° to 30° would increase the $V_{OC}$ from 1.30 to 1.34 V, which is mainly due to the reduction of the radiative recombination. It should be noted that the correlation between the restricted emission angle and $V_{OC}$ will become less significant if the SRH recombination is dominant as compared to the radiative recombination.

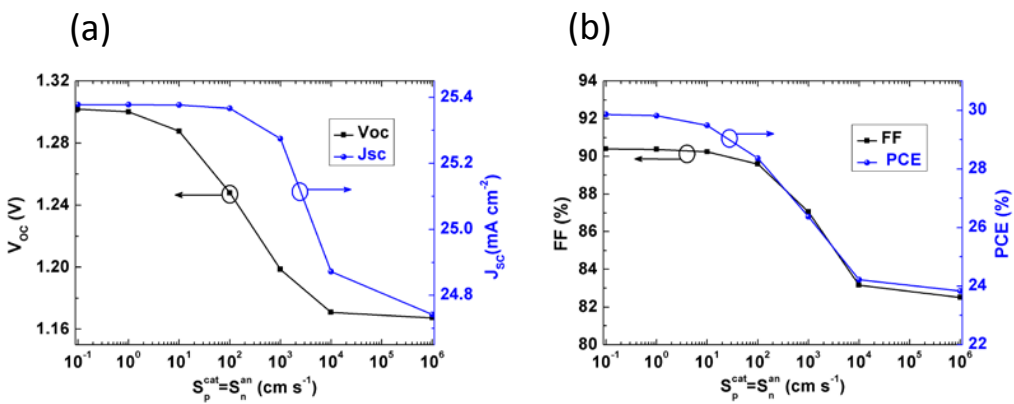

Figure S4. The electrical characteristics of the PVSCs with the nonselective contacts and different surface recombination velocities ($S_p^{cat} = S_n^{an}$ and $S_p^{an} = S_n^{cat} = \infty$). (a) $V_{OC}$ and $J_{SC}$, (b) FF and PCE.

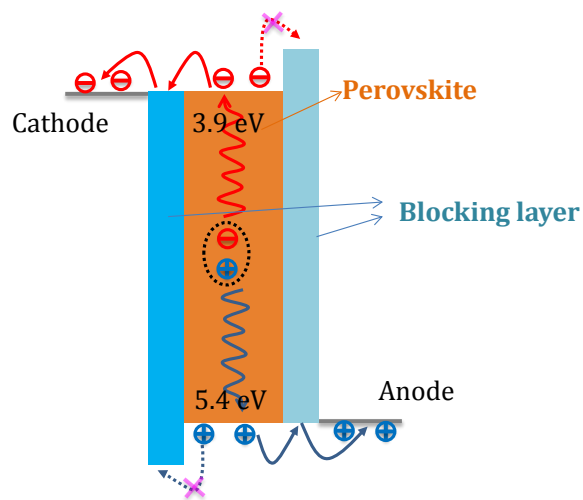

Figure S5. Schematic illustration of perovskite solar cells with the blocking layers. The blocking layers are inserted in between active layer and electrodes for eliminating the minority carrier recombination. The thickness of the blocking layers adopted in the simulation is 10 nm. For simplicity, mobility and dielectric constant of the blocking layer are set as the same as perovskite materials.

Table S2. The summary of PVSC performances with the different trap density $N_t$ and trap energy level $E_t$.

| Trap density $N_t$ ($10^{15}$ cm$^{-3}$) | | $V_{OC}$ (V) | $J_{SC}$ (mA cm$^{-2}$) | FF (%) | PCE (%) |
|---|---|---|---|---|---|
| $E_t$=0.05 eV | 100 | 1.23 | 25.38 | 89.61 | 27.89 |
| | 10 | 1.28 | 25.38 | 90.16 | 29.20 |
| | 1 | 1.30 | 25.38 | 90.35 | **29.76** |
| $E_t$=0.1 eV | 100 | 1.18 | 25.37 | 88.45 | 26.51 |
| | 10 | 1.24 | 25.38 | 89.67 | 28.22 |
| | 1 | 1.29 | 25.38 | 90.11 | **29.40** |
| $E_t$=0.15 eV | 100 | 1.14 | 25.34 | 85.36 | 24.58 |
| | 10 | 1.20 | 25.37 | 88.71 | 26.98 |
| | 1 | 1.26 | 25.38 | 89.43 | **28.58** |
| | 0* | 1.30 | 25.38 | 90.39 | 29.87 |

*$N_t$=0 cm$^{-3}$ represents the recombination only includes the bulk radiative recombination.

Regarding the variability/uncertainty of the physical parameter values of $CH_3NH_3PbI_3$ perovskite, we have conducted several simulations to demonstrate the variations of the device performance under the uncertainty of band gap ($E_g$), density of state ($N_C=N_V$), carrier mobility ($\mu_n=\mu_p$) and permittivity ($\varepsilon_r$), respectively.

Table S3. The summary of device performance with the varied values of the physical parameters for $CH_3NH_3PbI_3$ perovskite. The simulated solar cells are with the selective contact and no defect (i.e. for predicting the PVSC efficiency limit). The physical parameters of the control $CH_3NH_3PbI_3$ perovskite used in manuscript are $E_g$ =1.5 eV, $N_C=N_V=10^{21}$ cm$^{-3}$, $\mu_n= \mu_p$ =10 and $\varepsilon_r$ =10.

| **Selective contact +Defect free** | $V_{OC}$ (V) | $J_{SC}$ (mA cm$^{-2}$) | FF (%) | PCE (%) |
|---|---|---|---|---|
| **Control** | **1.30** | **25.38** | **90.39** | **29.87** |
| $E_g$+0.05eV | 1.30 | 25.38 | 90.39 | 29.87 |
| $N_C(=N_V)\times 10$ | 1.30 | 25.38 | 90.40 | 29.87 |
| $N_C(=N_V)\times 0.1$ | 1.30 | 25.38 | 90.39 | 29.87 |
| $\mu_n(=\mu_p)\times 10$ | 1.30 | 25.38 | 90.39 | 29.87 |
| $\mu_n(=\mu_p)\times 0.1$ | 1.30 | 25.38 | 90.39 | 29.87 |
| $\varepsilon_r\times 10$ | 1.30 | 25.38 | 90.39 | 29.87 |
| $\varepsilon_r\times 0.1$ | 1.30 | 25.38 | 90.28 | 29.83 |

In this case, the device configurations are suitable for the prediction of efficiency limit because there is no extrinsic loss in device modeling. In principle, the solar cell performance is only limited by the value of the radiative recombination in this device configuration. In our simulations, the variations of the physical parameters reveal no clear influence on the device performance. Therefore, the prediction capability of the advanced drift-diffusion model is very robust even there are some uncertainties of the physical parameters value (e.g. bandgap, mobility, density of state and permittivity) for $CH_3NH_3PbI_3$ perovskite.